\documentclass[sigconf, cidr]{acmart-cidr2021}

\AtBeginDocument{%
  \providecommand\BibTeX{{%
    \normalfont B\kern-0.5em{\scshape i\kern-0.25em b}\kern-0.8em\TeX}}}
    
\usepackage{xspace}
\usepackage{wrapfig}
\usepackage{subfigure}
\usepackage{boxedminipage}

\graphicspath{{./figures/}}

\newcommand{\eg}{e.g.,\xspace}
\newcommand{\ie}{i.e.,\xspace}

\newcommand{\dataset}{data set\xspace}

\newcommand{\datasets}{data sets\xspace}

\renewcommand{\paragraph}[1]{\vspace{0cm}\noindent\textbf{#1.}\xspace}

\newcommand{\hidecomment}[1]{}

\acmConference[CIDR '21]{11th Annual Conference on Innovative Data Systems Research, January 11-15, 2020, Chaminade, USA.}{January 11-15, 2021}{Chaminade, USA}

\begin{document}

\title{The Case for Distance-Bounded Spatial Approximations}

\author{Eleni Tzirita Zacharatou}
\affiliation{\institution{TU Berlin}}
\email{eleni.tziritazacharatou@tu-berlin.de}

\author{Andreas Kipf}
\affiliation{\institution{MIT CSAIL}}
\email{kipf@mit.edu}

\author{Ibrahim Sabek}
\affiliation{\institution{MIT CSAIL}}
\email{sabek@mit.edu}

\author{Varun Pandey}
\affiliation{\institution{TU Munich}}
\email{pandey@in.tum.de}

\author{Harish Doraiswamy}
\affiliation{\institution{NYU}}
\email{harishd@nyu.edu}

\author{Volker Markl}
\affiliation{\institution{TU Berlin and DFKI GmbH}}
\email{volker.markl@tu-berlin.de}

\begin{abstract}
Spatial approximations have been traditionally used in spatial databases to accelerate the processing of complex geometric operations.
However, approximations are typically only used in a first filtering step to determine a set of candidate spatial objects that may fulfill the query condition.
To provide accurate results, the exact geometries of the candidate objects are tested against the query condition, which is typically an expensive operation.
Nevertheless, many emerging applications (\eg visualization tools) require interactive responses, while only needing approximate results.  
Besides, real-world geospatial data is inherently imprecise, which makes exact data processing unnecessary. 
Given the uncertainty associated with spatial data and the relaxed precision requirements of many applications, this vision paper advocates for approximate spatial data processing techniques that omit exact geometric tests and provide final answers solely on the basis of fine-grained approximations.
Thanks to recent hardware advances,
this vision can be realized today.
Furthermore, our approximate techniques employ a \emph{distance-based error bound}, \ie a bound on the maximum spatial distance between false or missing and exact results which is crucial for meaningful analyses.
This bound allows to control the precision of the approximation and trade accuracy for performance.
\end{abstract}

\maketitle

\section{Introduction}
There is an explosion in the amount of spatial data being generated and collected
today. 
Billions of GPS-enabled mobile devices, cars, social networks, satellites, sensors,
and many other sources produce spatial data constantly.
As a result of the ever-increasing data sizes and the computationally-intensive nature of spatial queries, it is hard to provide fast response times, which opposes the
interactivity requirements of exploratory applications.

On the bright side, users often do not need exact results.
They are instead satisfied with approximate answers, especially if these answers are accompanied by precision guarantees.
However, approximate spatial data processing has attracted limited attention~\cite{Belussi2013, rasterapprox, onlinespatialaggr, spatial-synopses, deep-sampling, deepspace}.
There are two different notions of approximation in spatial databases.
Synopsis-based techniques aim to accelerate spatial queries by evaluating them on small samples or models of the data~\cite{onlinespatialaggr, spatial-synopses, deep-sampling, deepspace}.
Existing techniques in this category are limited to certain types of queries (\ie range queries, selectivity estimation, k-means clustering and spatial partitioning).
On the other hand, most spatial querying techniques approximate individual spatial objects with simpler geometries such as rectangles or convex polygons to accelerate queries~\cite{brinkhoff93, rasterapprox}. 
Unlike synopsis-based techniques, geometric approximations support arbitrary spatial predicates.
Our work is related to the latter category, \ie spatial query processing based on approximations of individual objects, which is orthogonal to sampling techniques that reduce the number of objects to be processed. 

Notably, prior work does not give guarantees on the \emph{spatial distance} between false (or missing) and exact results.
Consequently, it is hard to interpret the provided approximate results, as the user has no information about how closely these results correspond to the particular region she is interested in. 
Guaranteeing \emph{distance-based error bounds} is thus crucial. 
These bounds should be controlled by the user, essentially allowing to trade off between query results accuracy and query execution time. %

\paragraph{Motivating Application: Visual Exploration of Mobility Data}
In an effort to enable urban planners to make data-driven decisions, in early 2017 Uber introduced Uber Movement, a visualization platform for the exploration of Uber rides\footnote{\url{https://www.uber.com/newsroom/introducing-uber-movement-2/}}.
The platform allows users to visualize data of interest at different resolutions over varying time periods.
Such visual analyses require interactivity, %
since high latency reduces the rate at which users make observations, draw generalizations, and generate hypotheses~\cite{liu-heer@tvcg2014}.
Furthermore, exact answers are not required, because visualizations are approximate in nature.
Moreover, users typically perform ``level-of-detail'' exploration.
They first look at a high level overview, and then zoom into regions of interest for further details~\cite{Shneiderman96}. 
Finally, there is usually uncertainty with respect to spatial coordinates, as GPS positions are typically accurate to within a 4.9~m radius~\cite{van2015world}. 
Similarly, geographical region boundaries are often fuzzy, in the sense that adjacent regions are separated by extended zones (\eg a street surface) rather than one-dimensional lines. 
As a result, these zones can be considered to be part of any of the adjacent regions. 
Overall, the interactivity expected from exploratory applications (visual or not), coupled with the inherent properties of spatial data, necessitate a paradigm shift towards spatial data processing techniques that have approximation at their core.  

\paragraph{Hardware Trends}
Spatial approximations have been widely used in spatial databases.
Recent hardware trends, however, indicate that the time has come to rethink their design and utility. 
Existing techniques typically use a two-step ``filter and refine'' strategy~\cite{spatial_database_systems} where approximations are only employed in a first filtering step that yields a candidate result set. 
The subsequent refinement step eliminates false matches by performing exact geometric tests. 
Most efforts in improving spatial query processing focus only on the first filtering step \cite{spatial_join_techniques, quasii, stitch, space_odyssey, SM17}.
In the past decades, the filtering step was in the critical path of the execution, since it was fetching the spatial approximations from disk. 
As a consequence, the number of slow disk accesses had to be minimized, which led to the design of approximations that sacrifice precision for compactness.
Today's machines, however, have large DRAM sizes that can go up to multiple terabytes and are often equipped with large-capacity Non-Volatile Memory (NVM), making it unnecessary to store the approximations in slow secondary storage devices.
As a result of the faster access times provided by DRAM and NVM, the filtering step is no longer in the critical path, and the CPU-intensive refinement step becomes the bottleneck. 
As recent work shows~\cite{STIG2016}, the main memory-based filtering step takes only a few milliseconds even for billions of points.
Therefore, to improve performance, we now need to reduce (or even completely eliminate) the number of costly CPU-based refinements rather than the number of memory accesses.
This necessitates the re-design of spatial approximations: we can now easily store more precise (and thus larger) approximations and leverage fast random access storage in exchange for better filtering efficacy and fewer CPU-intensive operations.

With increasing data sizes, the computation of precise approximations becomes more expensive.
However, to support exploratory applications where the workload
changes dynamically, we need to compute spatial approximations fast and on-the-fly.
GPUs, and in particular their native support for \emph{rasterization} make that possible today.
The rasterization operation takes as input a geometric primitive (\eg a polygon) and converts it into a collection of pixels which essentially form a fine-grained uniform grid approximation of the primitive.
GPUs perform rasterization at interactive speeds, as they employ highly optimized hardware implementations.
This enables us to design techniques that leverage GPUs to compute spatial approximations and evaluate spatial queries in real time.  

This vision paper argues that fine-grained grid approximations can form the basis of spatial data processing.
We show that these approximations allow to provide distance-based error bounds, enable us to exploit modern hardware, and facilitate further optimizations such as the use of learned indexes. 
The remainder of this paper outlines our vision of incorporating distance-bounded spatial approximations in different components of a spatial system, highlights individual challenges, and presents promising initial results.  

\section{Approximate Processing}
In this section, we first present geometric approximations commonly used in spatial data processing.
We then describe how we can quantify the error that these approximations introduce.
Finally, we discuss the benefits of integrating distance-bounded spatial approximations in different components of a spatial system.  

\subsection{Geometric Approximations}
Spatial objects can have an arbitrarily complex structure. 
Even worse, different spatial objects can have very different structures (\eg a point is different from a polygon). 
To address this challenge, spatial query processing algorithms perform geometric tests (\eg intersection, containment) on approximations of the geometries~\cite{brinkhoff93}.
The employed approximations can represent objects with different geometries and retain the objects' main features.
In addition, they have a significantly simpler structure than the actual objects, which reduces computation and storage costs.

\begin{figure}
\includegraphics[width=.8\linewidth]{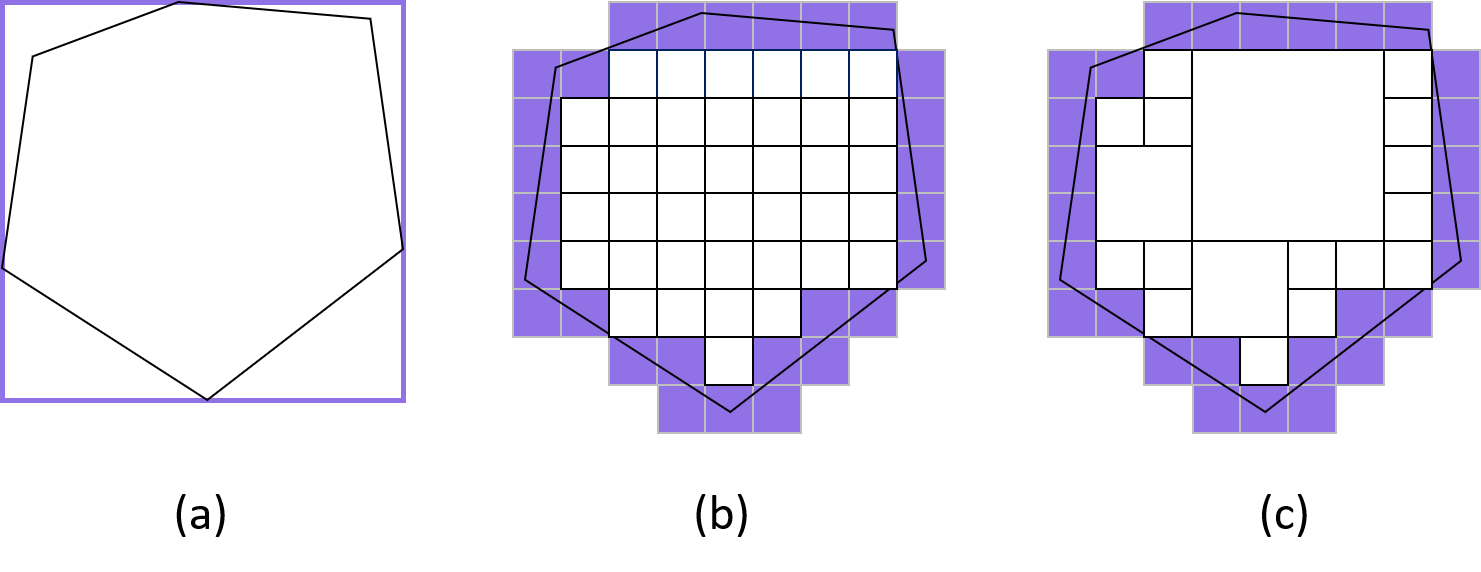}
\caption{Three example geometric approximations of a polygon: (a) Minimum Bounding Rectangle (MBR), (b) Uniform Raster (UR), (c) Hierarchical Raster (HR).}
\label{fig:approximations}
\end{figure}

The most widely used spatial object approximation is the Minimum Bounding Rectangle (MBR), which is the smallest axis-aligned rectangle that encloses the complete geometry of an object (Figure~\ref{fig:approximations}(a)). 
MBRs are rather rough and inaccurate approximations.
Clipped Bounding Rectangles~\cite{ltree} improve the accuracy of MBRs by clipping away empty space that is concentrated around the MBR corners. 
Brinkhoff et. al.~\cite{brinkhoff93} performed a detailed study of different approximations, namely the Rotated Minimum Bounding Rectangle (RMBR), the Minimum Bounding Circle (MBC), the Minimum Bounding Ellipse (MBE), the Convex Hull (CH), and the Minimum Bounding n-Corner (n-C).  
Raster approximations are another class of approximations that have recently attracted attention as they can provide high approximation accuracy. 
Raster approximations represent geometric primitives using a set of cells that can be either equi-sized~\cite{rasterapprox, raster-join} (Uniform Raster, Figure~\ref{fig:approximations}(b)) or variable-sized~\cite{kipf2020adaptive, geoblocks} (Hierarchical Raster, Figure~\ref{fig:approximations}(c)). 

Executing spatial queries on geometric approximations leads to approximate results that are typically further processed to obtain exact answers. 
However, when the geometric approximation is sufficiently precise and exact answers are not required, approximate query processing techniques can provide final answers solely on the basis of the approximate geometries. 
In this paper, we advocate for approximate techniques with application-driven accuracy and discuss next how to bound the approximation error.

\subsection{Distance Bound}
\label{sec:distance_bound}
\begin{wrapfigure}{r}{0.3\linewidth}
\centering
\includegraphics[width=\linewidth]{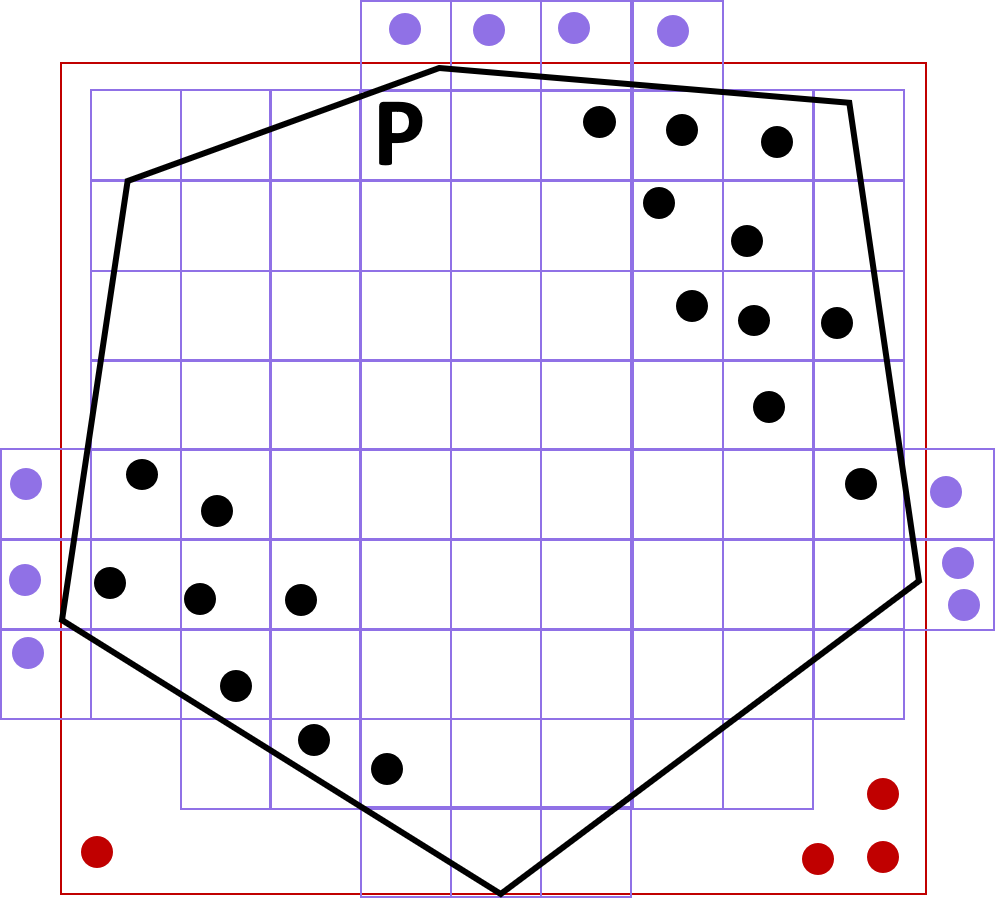}
\caption{Example polygon and points and two approximations of the polygon, MBR (red), and Uniform Raster (violet).}
\label{fig:toy-example}
\end{wrapfigure}

Spatial queries involve predicates that evaluate relations among objects in space (\eg intersection, containment).
Therefore, we argue that it is only natural for \emph{approximate} techniques to provide distance-based error bounds, \ie \emph{guarantees} on the spatial distance between false (or missing) and exact results. 
Approximate results without this notion of spatial distance can be misleading and hard to interpret. 
To illustrate this, consider the example in Figure~\ref{fig:toy-example}. 
It shows a set of points corresponding to the pickup location (latitude/longitude) of taxi rides. 
To optimize its operational planning, the taxi service provider needs to compute the count of trips that originate from within a given region $P$ depicted in the figure.
The exact count of taxis is 18. Consider now two approximate results.
The first one is computed over the set of black and red points and equals to 22, while the second one is computed over the set of black and violet points and equals to 28.
Although the first aggregate result is closer to the exact value, it contains points which are quite far away from the region $P$ that the user is interested in, while it does not include the violet points that are closer to $P$. 
We argue that for such exploratory analyses, the second result is more meaningful as it matches more closely the user's region of interest.
We further argue that in order to interpret the obtained approximate result, the user needs information about the spatial distance between the data points from which the approximate result was derived and the query geometry.
In other words, it is often admissible for the user to compute the result over a region that closely approximates $P$, as long as she knows how close in space the approximation is.

Formally, a geometry $g'$ $\epsilon$-approximates a geometry $g$ if the Hausdorff distance $d_H(g,g')$ between the two geometries is at most $\epsilon$,
where
\[ d_H(g,g') = \max \left\lbrace \max_{p' \in g'} \min_{p \in g} d(p,p'), \max_{p \in g} \min_{p' \in g'} d(p',p) \right\rbrace  \] 
and  $d(p',p)$ denotes the Euclidean distance between two points.
Intuitively, this ensures that any false positive (false negative) results that are present (absent) when answering queries using the approximate geometry $g'$ are within a distance $\epsilon$ from the boundaries of the original geometry $g$.

Interestingly enough, not all geometric approximations can be distance-bounded. 
The Hausdorff distance between an object and its MBR approximation is data dependent: the coordinates of the MBR corner points are the dimension-wise maxima/minima of the bounded object.
Consequently, the distance between a corner and the closest point in the object boundary can be very large. 

Raster approximations, in contrast, can be distance-bounded.
Given $\epsilon$, raster approximations such as the ones shown in Figure~\ref{fig:approximations}, can guarantee that $d_H(g,g')  \leq \epsilon$ by using a cell side length equal to $\epsilon' = \frac{\epsilon}{\sqrt{2}}$ (\ie the length of the diagonal of the cell is $\epsilon$) for the cells that are at the boundary of the geometry (shown with violet color). 
The interior cells that are fully contained in the geometry can have a cell side length larger than $\epsilon'$ as they do not contribute to the approximation error. 
 At the boundary, there can be two types of errors, depending on the implementation. 
If all the cells that overlap even the slightest with the boundary are part of the approximation, then there can only be \emph{false positive} results as the whole cells are considered to be part of the object.
We call such a raster approximation \emph{conservative}. 
In non-conservative raster approximations, the cells that have a small overlap with the boundary can be omitted, which can introduce \emph{false negative} results. 
Overall, the precision of raster approximations is independent of the geometry they approximate and \emph{tunable}.
This property makes them particularly suitable to form the basis of approximate spatial query processing techniques.

\subsection{The Power of Distance-Bounded Raster Approximations}
\begin{figure}
\includegraphics[width=.9\linewidth]{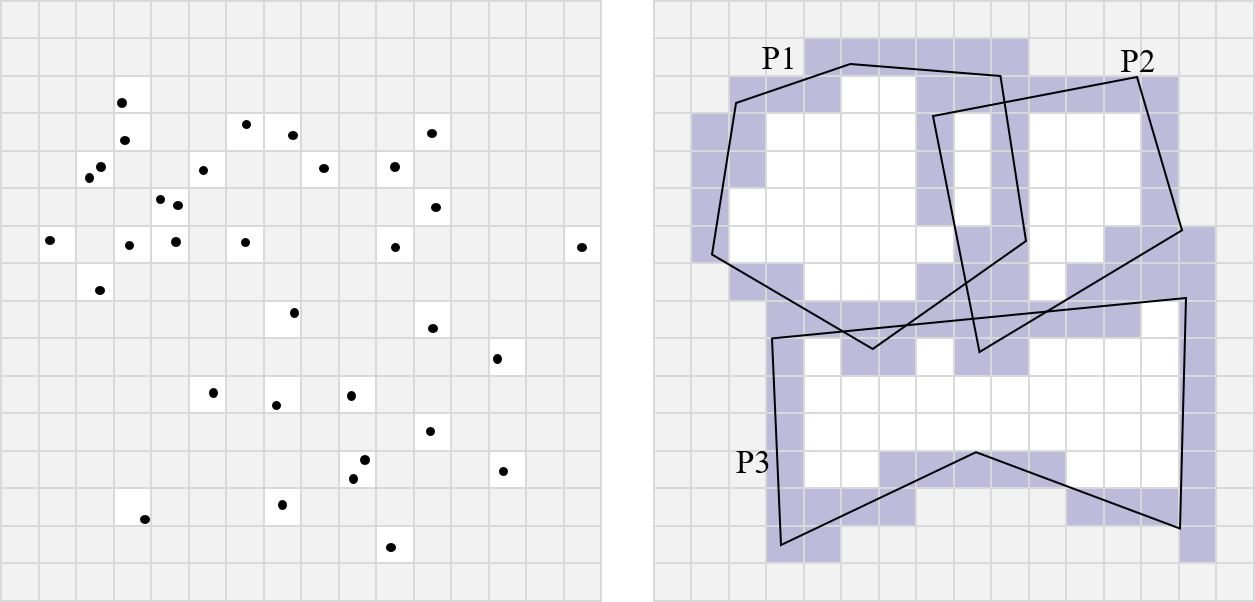}
\caption{Uniform Raster approximation of points (left) and polygons (right). Figure from \cite{raster-join}.}
\label{fig:representation}
\end{figure}

To illustrate the power of distance-bounded raster approximations, consider the example in Figure~\ref{fig:representation} showing two input \datasets, a set of points (left) and a set of polygons (right) approximated with UR.

\paragraph{Indexing} Figure~\ref{fig:representation} essentially shows how the data is represented \emph{logically}: geometric objects are approximated by a set of cells, potentially along with additional information that denotes the cells that intersect with the geometry boundaries. 
Given this representation, a database system needs efficient indexes to store the approximations and enable their fast retrieval. 
Since approximate query processing eliminates the expensive refinement step, the index lookup performance is crucial because it determines the query performance.
Traditional R-tree-based indexes~\cite{rst90} are not applicable as they are designed to index MBRs and are not compatible with raster approximations.  
At the same time, raster approximations enable new opportunities for a new generation of indexes.
Specifically, mapping the cells to a one-dimensional array by enumerating them with a space-filling curve, enables the use of a \emph{learned index}~\cite{radixspline2}.
As we show in Section~\ref{sec:data_access}, by learning the position of the cells in the 1D array, the learned index outperforms other spatial index structures.

\paragraph{Optimization} Section~\ref{sec:query_optimization} discusses how, by abstracting away from the specific object geometries and providing a \emph{unified} representation for different geometric data types, the raster approximation creates new opportunities in spatial query optimization.
That is, the implementation of primitive operations (\eg intersection tests) on the raster approximation can be \emph{independent} of the geometries and thus re-usable, while it can also leverage modern GPUs. 

\paragraph{Execution}
Other than enabling efficient access to a single \dataset,
the raster approximation also enables the efficient execution of queries that involve multiple \datasets, such as joins. As we show in Section~\ref{sec:query_execution}, by mapping geometries to sets of cells, we can observe the overlap at the cell level instead of performing geometry-to-geometry comparisons.
Each cell can be processed independently, which makes the computation highly parallelizable.
Furthermore, aggregations that are distributive or algebraic can be computed very efficiently.
The final aggregate can be obtained by combining partial aggregates calculated (in parallel) for each cell.

In the following, we describe how to use distance-bounded raster approximations in various system components in more detail and present initial results. 

\section{Data Access}
\label{sec:data_access}
Storage layouts and index structures determine the efficiency of data access. 
This section shows the details of how we can build high-performing indexes for polygon and point geometries that leverage raster approximations. 

\paragraph{Dimensionality Reduction}
While raster cells could be indexed using spatial data structures such as a Quadtree, a \emph{linearization} step can simplify the indexing problem significantly.
A common approach is to map 2D cells into a 1D domain by enumerating them with a space-filling curve, such as the Hilbert or Z curve.
As we will show, we can achieve much higher lookup performance with linearized cells, even compared to well-tuned 2D spatial indexes~\cite{learnedspatial}.

\paragraph{Polygon Indexing}
Given the \emph{logical} representation of polygons as a collection of linearized hierarchical cells, a database system can use different \emph{physical} representations to store these cells, such as a B+-tree or a sorted array.
Adaptive Cell Trie (ACT)~\cite{kipf2020adaptive,kipf2018approx} is a recently proposed radix tree data structure for indexing linearized cells of hierarchical raster approximations. 
A radix tree has a clear advantage over a B+-tree or a sorted array in this setting.
That is, matching cells can be found in any level of the tree, and larger cells are indexed closer to the root.
Hence, larger cells are likely to be found sooner during the tree traversal.
In addition, the radix tree offers implicit prefix compression as keys are not stored explicitly.

To index a set of polygons in ACT, we first perform a hierarchical raster approximation of the polygons that conforms to a user-defined distance bound (Section \ref{sec:distance_bound}). 
ACT uses the IDs of the linearized cells to build the radix tree.
To find a matching polygon for a query point, we first transform the query point to a cell on the most fine-grained grid level.
Then, we traverse the radix tree with the query cell of this point and retrieve the ID of the matching polygon (if such a polygon exists).

\paragraph{Point Indexing}
Like polygons, points are traditionally indexed with spatial data structures such as R-trees.
Here, we propose to apply the same linearization for mapping 2D points to 1D cell identifiers.
This again simplifies the indexing problem potentially leading to large speedups as we will demonstrate.
We store the resulting 1D cell identifiers (corresponding to 2D points) in a data structure such as a B+-tree or simply in a \textit{sorted} array.

To query the points with a polygon, we first approximate the query polygon using a hierarchical raster approximation, which yields a set of non-overlapping variable-sized cells that we call \textit{query cells}. 
Then, for each cell, we perform a binary search on the sorted array to get the qualifying points. For aggregation queries (\eg COUNT, SUM), one can pre-compute a prefix sum array and simply perform a lower and an upper bound lookup with the query cell's boundaries~\cite{prefixsum}.
By subtracting the lower bound from the upper bound, we can compute the aggregate value.
In this setting, the time for computing both lower and upper bounds (essentially a binary search each) really matters.
Therefore, we also explore using a learned index to speed up these searches.

We employ RadixSpline (RS) as a learned index~\cite{radixspline2}. RS consists of two main components: i) a set of spline points, and ii) a radix table to quickly determine the spline points to be examined for a lookup key (\ie the query cell in our case). 
At lookup time, we first consult the radix table to determine an initial range of spline points. 
Next, this range is searched over to determine the spline points surrounding the lookup key.
Finally, we use linear interpolation to predict the position of the lookup key in the sorted array.
Building RS requires only one pass over the data, and is thus efficient.

\begin{figure}
\subfigcapskip = -0.15cm
\centering
\subfigure[]{\label{fig:rs-latency}\includegraphics[width=.49\linewidth]{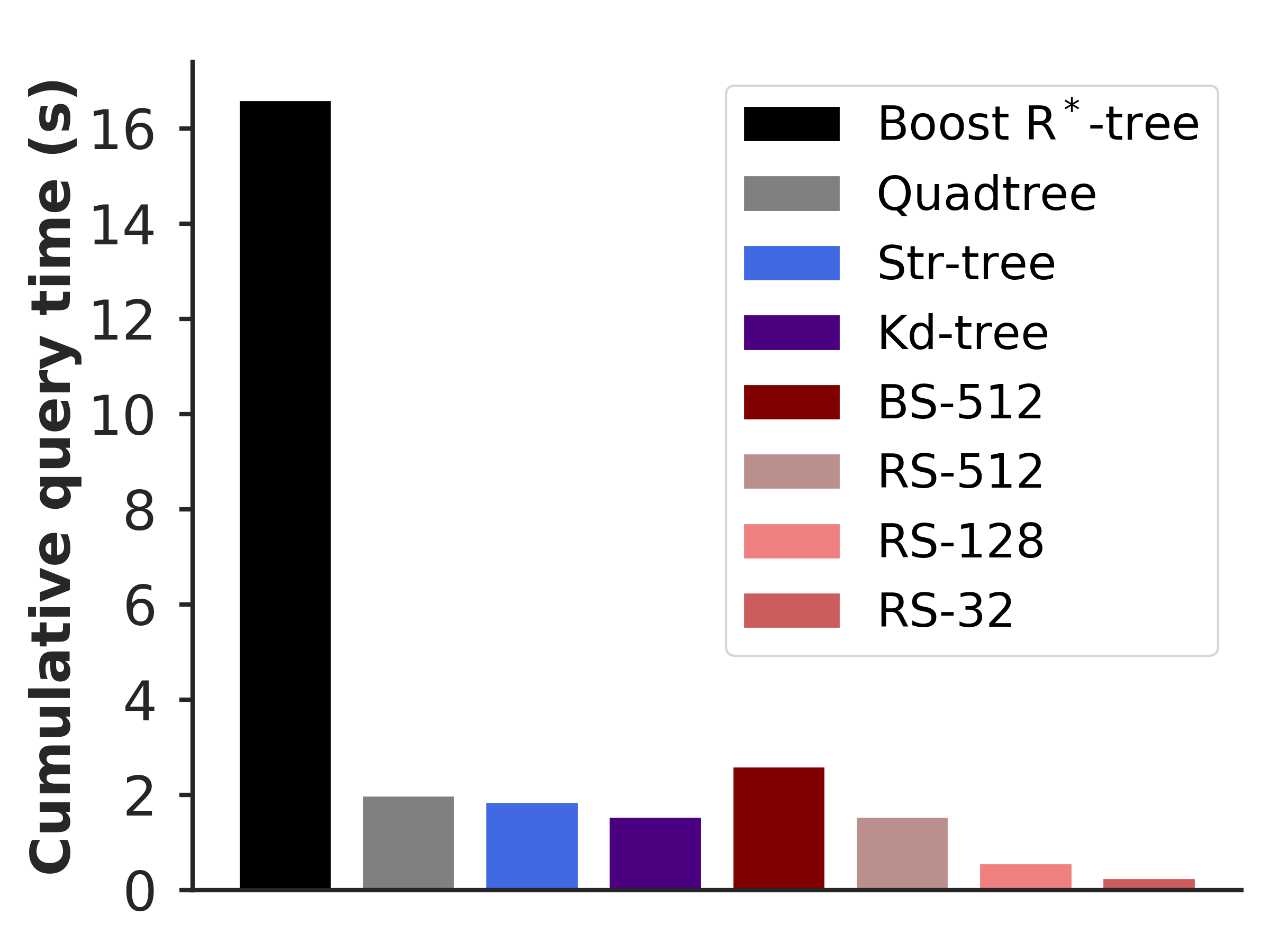}} 
\subfigure[]{\label{fig:rs-points}\includegraphics[width=.49\linewidth]{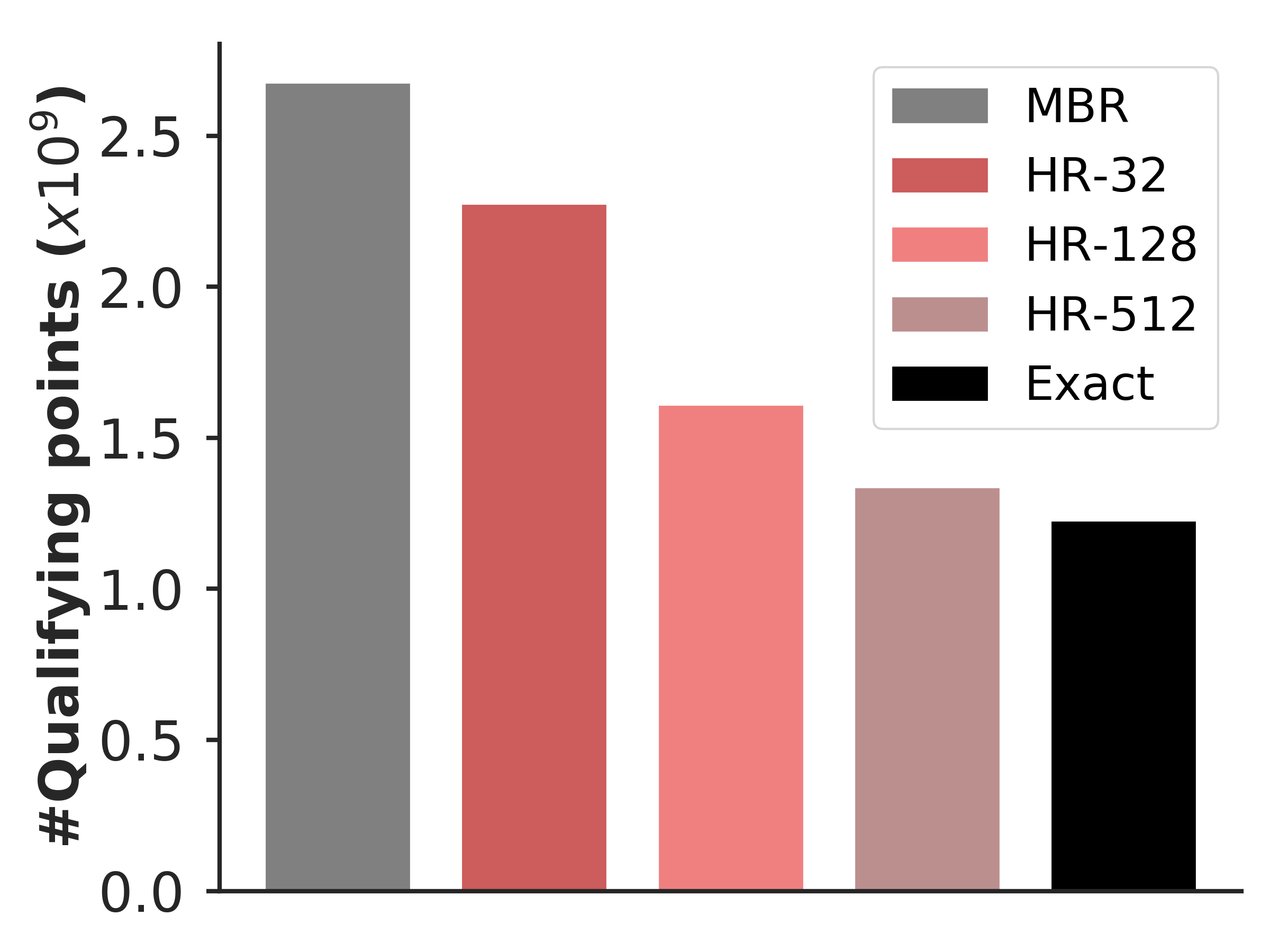}}
\caption{Data access efficiency. (a)~Point-polygon containment query performance. (b)~Impact of the precision of the raster approximation on the number of qualifying points.}  
\vspace{-2mm}
\label{fig:rs}
\end{figure}

\paragraph{Performance} We experimentally compare the performance of our proposed RS-based index with binary search (BS) and other four spatial indexes, namely, R$^\ast$-tree~\cite{rst90} from Boost Geometry~\cite{boostgeometry}, Quadtree~\cite{quadtree}, STR-packed R-tree~\cite{str_packing}, and Kd-tree~\cite{kdtree}. The spatial indexes act as baselines for filtering based on the MBR approximation. 
In our experiment, we use 39,200 polygons corresponding to the NYC Census regions (query polygons) and 1.2B points from the NYC taxi \dataset (years 2009 to 2016)~\cite{taxi-data}.
We implemented the Quadtree, the STR-packed R-tree, and the Kd-tree baselines based on recent research~\cite{learnedspatial}.
For the Boost R$^\ast$-tree, we chose the bulk-loading mode and manually optimized the number of elements per node. 
For the RadixSpline, we have set the number of radix bits to 25 and the spline error to 32.
This experiment was run single-threaded on a two-socket Arch Linux 5.7.4 machine with an Intel Xeon Gold 6230 Processor CPU (2.10\,GHz, 10 cores, 3.90\,GHz turbo)
and 256\,GB DDR3 RAM.

Figure~\ref{fig:rs}(a) shows the cumulative query time to find the total number of points inside the query polygons, while varying the precision of the raster approximation (\ie number of approximating cells per query polygon). We compared the results of three RS-based index variations, corresponding to three precision levels (32, 128, and 512 cells per polygon), with binary search at the highest precision level used (i.e., 512) and the other four spatial baselines. Note that the spatial baselines use MBR filtering, and hence they are agnostic to changing the precision level. Clearly, the three RS-based variations outperform both Boost R$^\ast$-tree and BS baselines (at least $10\times$ and 35\% better than Boost R$^\ast$-tree and BS, respectively). For the Quadtree, STR-packed R-tree, and Kd-tree baselines, the RS-based variations are still either better or very close to them in terms of query time. However, as shown in Figure~\ref{fig:rs}(b), RS-based variations are significantly better in terms of finding the tightest number of qualifying points compared to the exact number (precision level of 512 is almost similar to the exact case). Thus, in summary, our proposed RS-based index hits a sweet spot in the trade-off between precision and query time compared to all other baselines.

\section{Query Optimization}
\label{sec:query_optimization}

Existing approaches for spatial query processing are tied to specific geometric data representations and closely follow the relational model for query optimization~\cite{Samet1995}. 
They use operators that are tightly coupled to specific geometric types and query classes.
Let us consider again the selection query from Figure~\ref{fig:toy-example}. As mentioned earlier, this query is typically implemented as a \emph{single} operator that uses two phases: filtering and refinement. 
While the filtering phase relies on MBRs and is thus generic, the refinement phase depends on the geometric type and operation. In this example, the refinement is specific to the input being points, and the performed operation is a point-in-polygon test.
If the input changes from taxi pickup locations to restaurants represented by polygons, then a different implementation is required, since a polygon-intersect-polygon test must be performed instead.
The use of such large monolithic operators limits the set of options over which optimization can be performed, as the operators cannot be reused across query classes. 

\begin{figure}
\includegraphics[width=.9\linewidth]{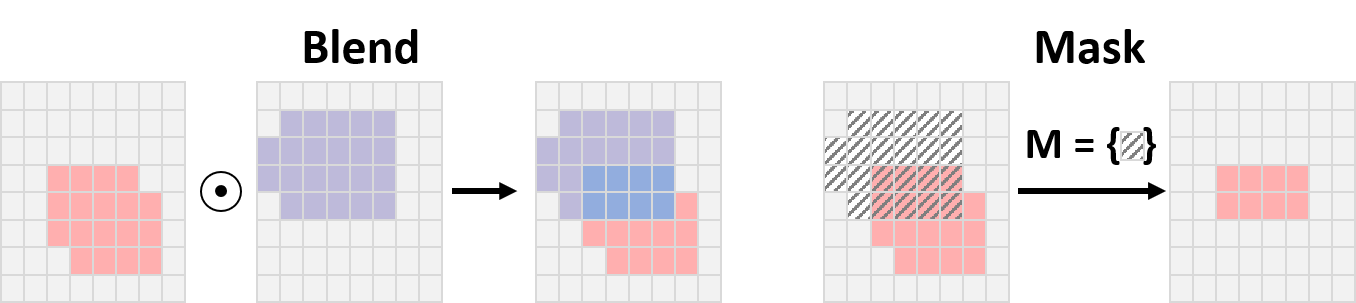}
\caption{The blend and mask operators applied on rasterized canvases. The different colors are used for illustrative purposes to denote the information stored in each pixel of the rasterized canvas; the grey color denotes empty pixels.}
\label{fig:operators}
\end{figure}

To overcome these limitations, and to exploit modern GPUs, a GPU-friendly spatial data model and algebra was introduced in \cite{spatial_model_extended}, which proposes a uniform data representation called \textit{canvas} and a small set of simple parallelizable operators.
These operators include common computer graphics operations: \textit{blend}, \textit{mask}, and \textit{affine transformations}.
More importantly, these operators are sufficient to realize common spatial query classes without being tied to specific geometries. 
For instance, both point-polygon and polygon-polygon intersection tests boil down to applying a combination of the above operations on the canvas. 
We propose to adapt the canvas model to support distance-bounded approximate queries: the canvas now simply becomes a rasterized image, where the pixel size depends on the required bound. The GPU-amenable operators work directly on such a \textit{rasterized canvas}---in fact, the implementation of these operators now becomes straightforward since boundary conditions~\cite{spatial_model_extended} need not to be taken care of. 
Figure~\ref{fig:operators} illustrates examples for the blend and the mask operators.
The \textit{blend} binary operator merges two rasterized canvases into one. 
The blend function $\odot$ defines how the merge is performed. 
The \textit{mask} operator filters pixels of the rasterized canvas to retain only those pixels that satisfy the condition specified by $M$.
There are two ways to generate a rasterized canvas: by rendering the data directly on the GPU, or through the use of indexes (\eg using ACT described in Section~\ref{sec:data_access}).

The rasterized canvas along with the proposed set of operators enable the creation of multiple alternative plans to realize any given ad-hoc query, thereby adding flexibility in the optimization process.
Furthermore, each operator can have multiple implementations and indexes can be reused across operators, which provides a wider set of options for the optimizer.
Thus, the optimizer can choose different query plans based on the query parameters, the distance bound (\ie the resolution of the rasterized canvas), and the estimated selectivity.
As an example of the potential gains that our proposed model provides, we show in Section~\ref{sec:raster_join} how the model allows for an alternate plan for an approximate spatial aggregation query that performs significantly faster than traditional approaches.

\section{Query Execution}
\label{sec:query_execution}

This section highlights the benefits of distance-bounded raster approximations in query evaluation.
As a representative example, we focus on the evaluation of spatial aggregation queries defined as follows in SQL-like notation:

\noindent 
\begin{boxedminipage}[b]{\linewidth}{}
\small
\texttt{SELECT AGG($a_i$) FROM P, R \\
WHERE P.loc INSIDE R.geometry [AND filterCondition]*\\
GROUP BY R.id}
\end{boxedminipage}

\noindent Given a set of points of the form $P(loc, a_1, a_2, \dots)$,
where $loc$ and $a_i$  are the location and attributes of the point,  
and a set of regions $R(id, geometry)$, this query performs 
an aggregation (\texttt{AGG}) over the result of the join 
between $P$ and $R$.
The geometry of a region can be any \emph{arbitrary polygon}. 
Functions such as COUNT($*$) or AVG($a_i$) are commonly used for \texttt{AGG}.

This query typically uses point-in-polygon (PIP) tests to identify polygons that contain each of the points. Note that each PIP test requires time linear with respect to the size of the polygon.
Since real-world polygonal regions often consist of hundreds of vertices, these tests are computationally intensive.
This challenge is compounded by the fact that \datasets can have hundreds of millions, or even billions of points, requiring a large number of PIP tests to be performed.

Existing systems typically evaluate spatial aggregation queries by performing a spatial join of the points and the polygons, followed by the aggregation of the join results.
To reduce the number of PIP tests, the join is first solved using MBR approximations.
As we show next, our evaluation strategies that are based on raster approximations, outperform the above approach significantly. 

\subsection{Main-Memory Join}
 Using our ACT index (Section~\ref{sec:data_access}), we can evaluate the query with an index-nested loop join: we simply index the polygons with ACT, and query the radix tree for every point.
 We combine the join with the aggregation to avoid materializing the join result.
 Given that ACT employs a fine-grained distance-bounded HR approximation, we omit the PIP tests and provide approximate results.
 
\begin{wrapfigure}{r}{0.5\linewidth}
\centering
\includegraphics[width=\linewidth]{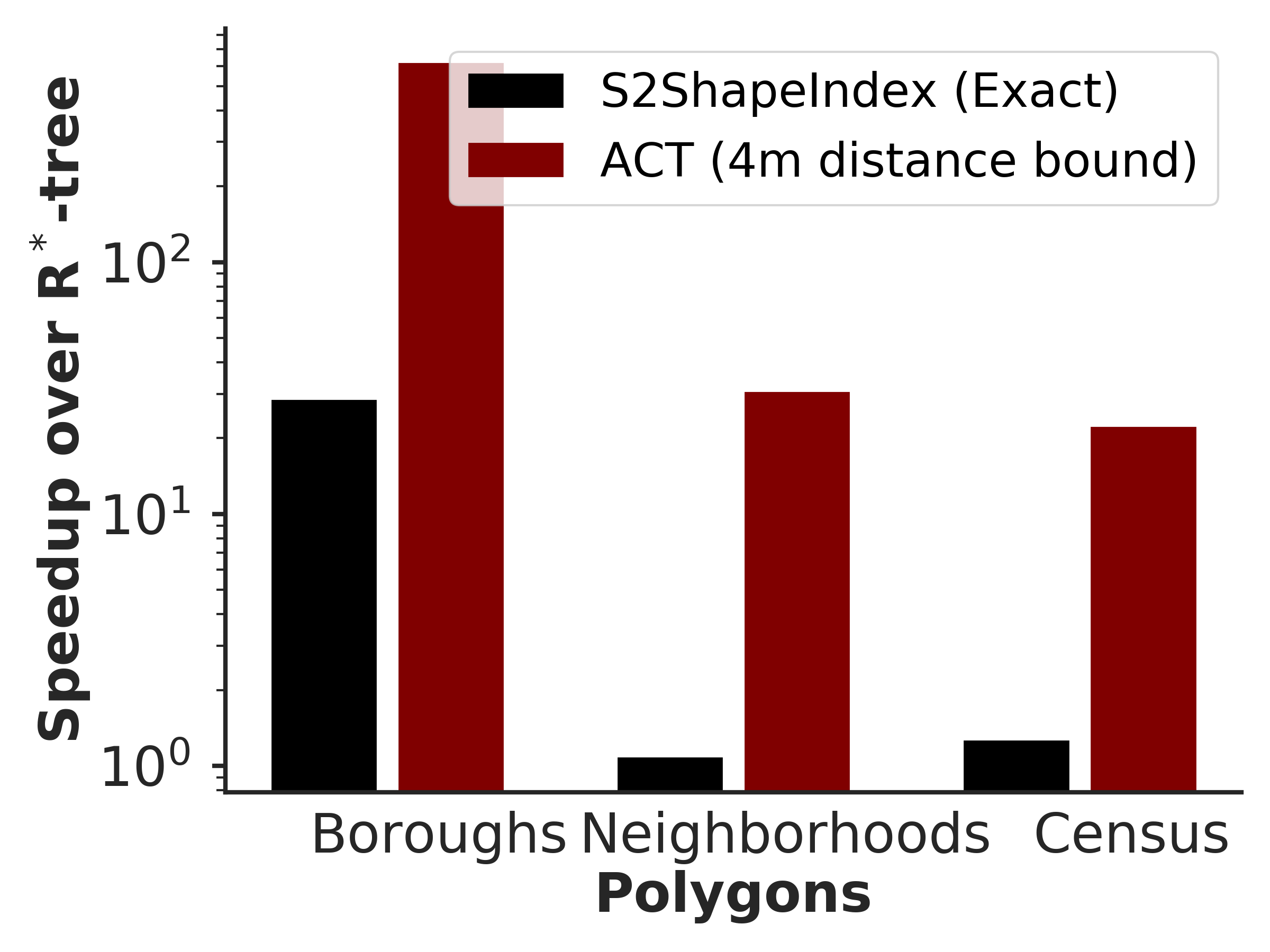}
\caption{Main-memory join.}
\label{fig:act}
\end{wrapfigure}

 \paragraph{Performance} We experimentally compare the performance of our approximate join with exact joins using the Boost~\cite{boostgeometry} R$^\ast$-tree~\cite{rst90} and Google's S2ShapeIndex (SI)\footnote{\url{https://s2geometry.io/devguide/s2shapeindex}}, all implemented in C++.  
ACT uses HR polygonal approximations satisfying a 4m distance bound. 
The R-tree indexes the polygons’ MBRs, while, similarly to ACT, SI uses HR approximations. 
However, SI's approximation is not distance-bounded and SI does not support approximate evaluation. 
We use 1.2B points from the NYC taxi \dataset~\cite{taxi-data} and three NYC polygon \datasets: Boroughs (5), Neighborhoods (289), and Census (39,200).
This experiment was run single-threaded on a machine with 14-core Intel Xeon E5-2680 v4 CPUs and 256\,GB DDR4 RAM.
Figure~\ref{fig:act} shows that the ACT-based approximate join significantly outperforms other approaches. Compared to the R$^\ast$-tree, it brings over two orders of magnitude improvement for Boroughs, and over one order 
otherwise, while it is over one order of magnitude faster than SI in all cases.
The low performance of the R$^\ast$-tree for Boroughs is due to the fact that Boroughs are more complex polygons than Neighborhoods and Census and thus PIP tests are more expensive.
Specifically, Boroughs have 663 vertices per polygon on average, while Neighborhoods have 30.6 and Census 13.6.
Therefore, reducing the number of PIP tests by approximating the polygons more closely (as SI does) or completely eliminating them by using distance-bounded fine-grained approximations (like ACT) has a significant impact on performance.
On the contrary, in the case of Census, which are the simplest polygons, ACT brings the least improvement.
Experiments on other \datasets~\cite{kipf2020adaptive} also confirm the above findings. 

Overall, ACT trades memory consumption for approximation accuracy, which in turn enables approximate evaluation and leads to higher performance.
Therefore, ACT has higher space consumption compared to the other approaches.
For example, the HR approximation of the Neighborhood polygons consists of 13.2M cells, which are represented using 64-bit integer IDs.
The total size of ACT is 143\,MB. 
In contrast, SI that uses a coarser-grained HR approximation occupies 1.2\,MB, while the R$^\ast$-tree that approximates polygons even more coarsely using MBRs occupies only 27.9\,KB.

\subsection{GPU Join}
\label{sec:raster_join}
Section~\ref{sec:query_optimization} outlined the use of a rasterized canvas model for executing spatial queries on GPUs. Here we show the gains that the proposed model brings in the evaluation of spatial aggregation queries. 
In fact, the query can be realized by simply combining a small set of operators from our query algebra on top of the rasterized canvas model.
This is exactly what our recently proposed algorithm, Bounded Raster Join~\cite{raster-join, raster-demo} (BRJ), does. 
Intuitively, BRJ takes as input a uniform representation of the points and polygons on rasterized canvases. 
It then merges (using the \textit{blend} operator) all the points into a single canvas that maintains partial aggregates, \ie each canvas pixel keeps the aggregate of all points falling in that pixel.
Then, it joins this canvas with the set of polygon canvases (by composing  the \textit{blend} and \textit{mask} operators) to identify points that intersect with the
polygons, and finally merges the results (using a combination of \textit{transformations} and \textit{blending}) to compute the final aggregates. 
That is, it combines the aggregates from the
individual pixels that fall within a polygon to generate the aggregation for that polygon.
The precise query plan can be found in~\cite{spatial_model_extended}. 
The above operations are natively supported by the graphics pipeline, leading to orders of magnitude speedup over typical evaluation strategies on CPUs without requiring any pre-computation~\cite{raster-join}.

\paragraph{Performance} 
We implemented BRJ using C++ and OpenGL.
We create the canvases on-the-fly by simply rendering the geometries onto an off-screen buffer and store the aggregates in the buffer's color channels (r,g,b,a). 
We experimentally compare BRJ with an accurate GPU Baseline that follows the traditional index-based evaluation strategy of first filtering the polygons with a grid index (with $1024^2$ cells) and then performing PIP tests. 
This experiment was run on a machine with an Intel Core i7 Quad-Core CPU, 16\,GB RAM, and an NVIDIA GTX 1060 mobile GPU with 6\,GB of memory, out of which we use only 3\,GB.
We join 600M points of the NYC taxi \dataset~\cite{taxi-data} (transferred in batches to the GPU) with 260 NYC neighborhood polygonal regions (some of the regions are multi-polygons) and count the number of points in each region.
\begin{wrapfigure}{r}{0.45\linewidth}
\centering
\includegraphics[width=\linewidth]{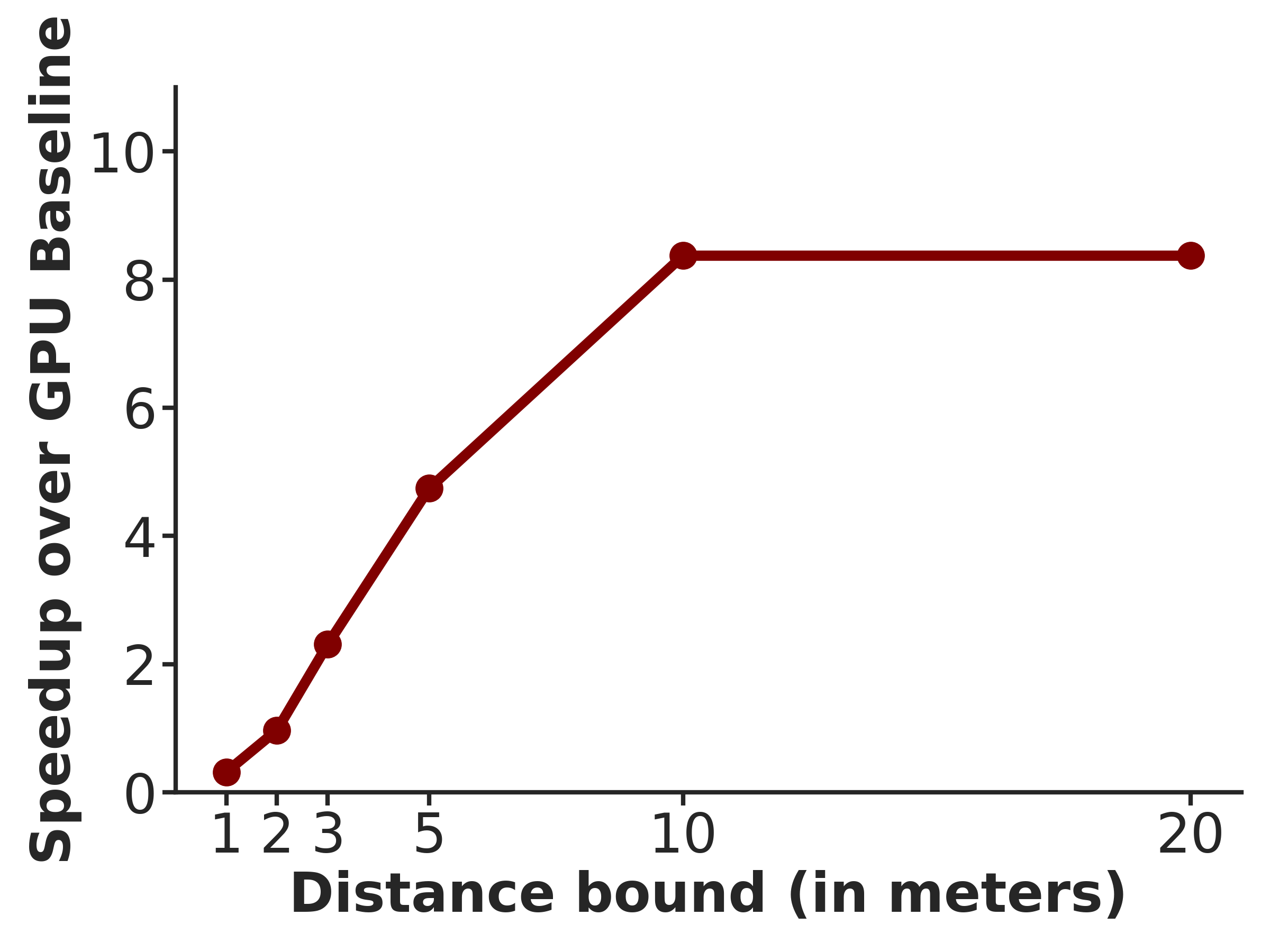}
\caption{Bounded Raster Join (GPU). Impact of the distance bound on performance.}
\label{fig:raster-join}
\end{wrapfigure}
Figure~\ref{fig:raster-join} shows that there is a trade-off between the accuracy and the query time.
For a distance bound of 10m, BRJ is about $8.5\times$ faster than the baseline, while for 1m it becomes slower.
This is because lower bounds require smaller pixel sizes, and hence increasing the canvas resolution. 
When this resolution becomes higher than what the GPU supports, BRJ needs to divide the rasterized canvas and perform multiple aggregations, one for each subdivision.  
We note, however, that with a distance bound of 10m we get close to accurate counts: over all the polygons, the median error is only about 0.15\%.
BRJ can therefore provide a significant speedup with only a small accuracy loss. 
The accuracy-time trade-off has a similar behavior for larger inputs as well as for other \datasets~\cite{raster-join}.

\section{Discussion}
\paragraph{Synopsis-based Approximate Spatial Query Processing} 
Approximate Query Processing (AQP) typically refers to extracting small data synopses (\eg samples) from large spatial \datasets, and performing accurate evaluation on top of those samples, yielding approximate answers due to the initial data reduction~\cite{spatial-synopses}.  
Prior work in that direction~\cite{onlinespatialaggr, deep-sampling}
does not provide support for arbitrary spatial queries such as joins and group-by predicates.
Furthermore, most existing methods do not provide any accuracy guarantees and do not have the notion of distance bounds.
Initial efforts to provide such guarantees~\cite{deep-sampling} focus on the selectivity estimation problem and only provide bounds on the relative error between the actual and the estimated selectivity.

The above line of work is orthogonal to what we propose in this paper. 
We focus on approximations in space, \ie approximations of individual object geometries, and on tunable distance bounds that control the spatial accuracy of the approximations.

\paragraph{Result Range Estimation}
Rather than providing only an approximate result, we can use the raster approximation to provide a result range based on the key insight that errors happen only at the boundary cells. 
Therefore, by counting the number of results contained in these cells we can get loose bounds on the result range.
For example, let us assume that we have a \emph{conservative} raster approximation, \ie we can only have false positives at the boundary, and let $\alpha$ be the approximate count of points within a polygon.  
Let $C$ be the set of cells at the boundary and $\epsilon$ be the partial count computed over $C$. 
Then, we know that the result falls in the interval [$\alpha - \epsilon,\alpha$] with 100\% confidence.
In the above calculation, we assume that all the results at the boundary are false positives, which is the worst case. 
By making some assumptions about the distribution of points at the boundary, we can obtain a tighter interval.

\paragraph{Higher-Dimensional Data}
Even though this paper focuses on 2D primitives, the proposed distance-bounded approximation can be directly extended to support 3D primitives.
However, the proposed operators do not have a straightforward GPU implementation over 3D data.
In our future work, we plan to investigate extensions to our techniques to handle 3D data.

\paragraph{GPU Rasterization vs. Ray Tracing} 
This work shows the benefits of using the GPU rasterization pipeline in spatial data processing. 
Given that spatial databases rely on the same primitive types (geometric objects) and operations that are similar to the ones used in graphics (\eg spatial selections), we expect further opportunities to exploit advanced graphics techniques and hardware in the design of spatial systems.
In future work, it will be particularly interesting to explore the use of native GPU ray tracing, recently introduced by RTX GPUs from Nvidia~\cite{rtx}.
Ray tracing can be, for example, used to support 3D spatial queries.

\balance
\section{Conclusion}
Changes in applications requirements and hardware have been the main driving forces in rethinking the role of geometric approximations in spatial data management. 
This paper shows that distance-bounded raster approximations can enable a wider set of optimization options and can form the basis of approximate spatial query processing techniques that take better advantage of modern hardware and improve performance.
Our experiments demonstrate that raster approximations can be indexed efficiently and can provide a sweet spot in the trade-off between precision and query time. 
In doing so, we set the stage for new spatial systems that employ distance-bounded raster approximations at their core.

\begin{acks}
This work was partially supported by the German Ministry for Education and Research as BIFOLD - Berlin Institute for the Foundations of Learning and Data (ref. 01IS18025A and ref 01IS18037A).
This research was further supported by Google, Intel, and Microsoft as part of the MIT Data Systems and AI Lab (DSAIL) at MIT, NSF IIS 1900933, DARPA Award 16-43-D3M-FP040.
Ibrahim Sabek was supported by the NSF, under grant \#2030859 to the Computing Research Association for the CIFellows Project.
Harish Doraiswamy was supported in part by the NYU Moore Sloan Data Science Environment and the NSF award CCF-1533564.
\end{acks}

\bibliographystyle{abbrv}
\bibliography{paper.bib} 

\end{document}